\documentclass[modern]{aastex631}

\usepackage{hyperref}
\let\tablenum\relax % Issue with siunitx/AASTeX
\usepackage{siunitx}
\usepackage{graphicx}%
\usepackage{multirow}%
\usepackage{amsmath,amssymb,amsfonts}%
\usepackage{amsthm}%
\usepackage{mathrsfs}%
\usepackage[title]{appendix}%
\usepackage{xcolor}%
\usepackage{textcomp}%
\usepackage{booktabs}%
\usepackage{algorithmicx}%
\usepackage{algpseudocode}%
\usepackage{listings}%
\usepackage{upgreek}%

\begin{document}

\title{LunaIcy: Exploring Europa's Icy Surface Microstructure through Multiphysics Simulations}

\correspondingauthor{Cyril Mergny}

\author[0009-0002-1910-6991]{Cyril Mergny}
\affiliation{Université Paris-Saclay, CNRS, GEOPS \\
Orsay, 91405, France}
\email{cyril.mergny@universite-paris-saclay.fr}

\author[0000-0002-2857-6621]{Frédéric Schmidt}
\affiliation{Université Paris-Saclay, CNRS, GEOPS \\
Orsay, 91405, France}
\affiliation{Institut Universitaire de France \\
Paris, France}

\begin{abstract}
% 250 words max    
A multiphysics simulation model incorporating a sintering model coupled with the MultIHeaTS thermal solver, was developed to study the evolution of icy moons' microstructure.
The sintering process is highly dependent on temperature, and this study represents the first attempt in planetary science to examine the coupled interaction between heat transfer and sintering.
Our approach to ice sintering is based upon the literature, while offering a refined description of the matter exchange between grains, bonds, and the pore space.
By running the numerical framework, we simulate the evolution of ice microstructure on Galilean satellites, specifically tracking the changes in the ice grain and bond radii over time.
LunaIcy, our multiphysics model, was applied to study the evolution of Europa's ice microstructure over a million years along its orbit, with a parameter exploration to investigate the diverse configurations of the icy surface.
Our results indicate that effective sintering can take place in regions where daily temperatures briefly surpass 115 K, even during short intervals of the day. 
Such sintering could not have been detected without the diurnal thermal coupling of LunaIcy, due to the cold daily mean temperature.
In these regions, sintering occurs within timescales shorter than Europa's ice crust age, suggesting that, in present times, their surface is made of an interconnected ice structure.

%Accurately simulating these highly coupled processes, can contribute to the refinement of surface measurements like spectroscopy, enabling improved constraints on grain size.

\end{abstract}


\section{Introduction}

The surfaces of icy moons have a microstructure shaped by a complex interplay of physical processes. 
Among them, ice \textit{sintering}, also known as  \textit{annealing} or \textit{metamorphism},  transports material from ice grains into their bond region, resulting in changes in the thermal, optical and mechanical properties of the ice \citep{Adams2001, Blackford2007}. 

On Earth, the sintering process is typically studied using ice samples of snow, firn or ice.
Using these ice core measurements, snow metamorphism is implemented in a phenomenological way through a set of quantitative laws describing the evolution rate of bond and grain radii.
The state of the ice or snow, whether it is dry or wet, if it possesses a temperature gradient or not, significantly influences the conditions of sintering, resulting in different types of rate equations \citep{Vionnet2012}. 
From these empirical laws, multiphysics snow cover models such as SNTHERM \citep{Jordan1991},  CROCUS \citep{Brun1992,  Vionnet2012} or  SNOWPACK \citep{Lehning2002b}, can simulate the evolution of the ice microstructure as a function of the environmental conditions.

While the sintering process of snow on Earth has been extensively studied, there is a scarce amount of information regarding the alteration of ice in planetary surface environments characterized by low temperatures and pressures.
In the planetary science community, the basis for the sintering equations originates from \citep{Swinkels1981}, where the sintering of grains is studied for metallurgy. These equations have been applied to the sintering process on comet surfaces \citep{Kossacki1994}, Mars \citep{Eluszkiewicz1993a} and the surfaces of icy moons \citep{Eluszkiewicz1991, Eluszkiewicz1993,Schaible2017, Gundlach2018, Molaro2019, Choukroun2020}.

The kinetics of sintering  are highly dependent on temperature \citep{Molaro2019}, so achieving a precise temperature estimation is essential for accurately characterizing this process. However, such investigation has not been conducted primarily due to a significant challenge:
the substantial difference between the surface temperature variations timescale and the sintering timescale under cold conditions.
In this study, we introduce a multiphysics model to investigate how thermal coupling influences the sintering process in simulations covering a timeframe of one million years.

While the sintering process and the methodology used here can be applied to any icy surface in the Solar System, our primary focus is to investigate the evolution of Europa's ice microstructure.
An estimation of the sintering timescale on Europa at constant temperature leads to a lower limit of $\SI{e4}{years}$  and upper limit of $\SI{e9}{years}$ \citep{Molaro2019}.
The authors suggested that, in reality, the daily variations of temperature would need to be taken into account to obtain the accurate sintering timescale falling between these two limits.
This aspect is particularly intriguing given Europa's estimated ice crust age of 30 million years \citep{Pappalardo1998} leading the question:
Does Europa's ice microstructure evolve on a timescale shorter than its ice crust age? If so, what are the conditions on specific parameters (albedo, grain size, latitude, ...) necessary for sintering? Answering these questions would provide valuable insights into the current state of Europa's ice microstructure.

Inspired by snow cover models used on Earth, we developed a first-generation multiphysics model to simulate ice microstructure on planetary surfaces. The following section is dedicated to the description of the sintering model, the third section to the multiphysics coupling, and results and discussion are presented in the fourth section.

%Understanding ice microstructure is crucial in the planetary science field as it can give records on past changes, such has surface displacements, creation age (cryomagma cooling), and/or pas thermal anomalies.
%It is also necessary to prepare for surface in situ space exploration, has different microstructure lead to different mechanical and thermal behavior of the icy surface which could be hazardous for rovers.
%Finally, models can be used in conjonction with surface data analysis that measure grain sizes, etc... \textcolor{red}{Cite Cruz-Mermy, Hansen, ...}
%Simulating the time and space evolution of the snowpack is key to many scientific and socio-economic applications, such as weather, hydrological (flood predictions and hy- dropower) and avalanche risk forecasting in snow-covered \citep{Vionnet2012}.

\section{Sintering Model}

Early studies on sintering \citep{Kingery1955, Hobbs1964, Swinkels1981, Maeno1983} have identified six mechanisms of diffusion for the sintering process.
In the conditions of low temperatures found on icy moons, the dominant mechanism is diffusion by vapor transport between the grain and bond surfaces via the pore space \citep{ Kossacki1994,Blackford2007, Gundlach2018,  Molaro2019}.
The convex grain surface sublimates matter into the pore space, which then condenses on the concave bond surface.
This diffusion of water vapor occurs as a result of the different curvatures of the grain and bond surfaces, which changes the local saturation pressure.
In order to quantify this process, we have developed a model to describe  the evolution of gas mass within the pore space over time.

The focus of this approach is to accurately describe the exchange of matter between the grain, the bond and the pore space. 
Previous models used in planetary science, directly compute the mass flux from the grain surface to the bond surface, as it was suggested by \cite{Swinkels1981}. 
However this difference of curvature between the grains and bonds, does not appear in any of the analytical models of sintering used on Earth \citep{Miller2003, Flin2003, Flanner2006}.
The evaporation-condensation laws  may have been incorrectly used for planetary cases, as they do not seem  to properly described the exchange of matter between the ice and the gas in the pore space.
To address this, in our model, we  calculate both mass fluxes separately, the flux of matter from the grain surface to the pore space and the flux between the pore space and the bond.

\begin{table}[htbp]
    \centering
    \begin{tabular}{l c l l}
    \toprule
    Parameter & Value  &  Unit & Reference \\ 
    \midrule
    Bulk ice density $\rho_{\mathrm{0}}$ & 917  & $\SI{}{kg.m^{-3}}$ & \citet{Molaro2019}  \\
    Ice heat capacity $c_{\mathrm{p}}$ & 839  & $\SI{}{J.kg^{-1}.K^{-1}}$ & \citet{Klinger1981}  \\
    Water surface tension  $\gamma$ & 0.06 & \SI{}{J.m^{-2}} & \citet{Molaro2019} \\
    Ice shear modulus $\mu$ & \SI{e9}{} & \SI{}{N.m^{-2}} & \citet{Molaro2019} \\
    \bottomrule
    \end{tabular}
\caption{Constant ice properties used in our model.}
\label{tab:properties}
\end{table}

\subsection{Model Geometry}

\begin{figure}[htpb]
	\centering
	\includegraphics[width=\textwidth]{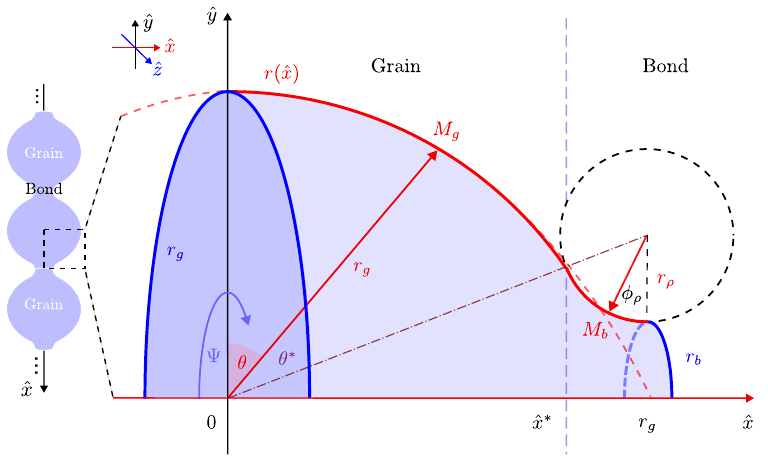}
    \caption{(\textit{Left}) Grain/bond pattern repeated periodically along the $\hat{x}$ axis. Each grain is connected to its neighbors by top and bottom bonds. (\textit{Right})  Schematic representation of the grain/bond geometry. Due to symmetry along two axes represented here, our model only needs to represent a quarter of total grain/bond volume.  The red line follows the curvature of the vapor/ice interface along the $\hat{x}$ axis, while the blue line follows the curvature in the $\hat{y}\hat{z}$ plane.}
    \label{fig:meta_geom}
\end{figure}

The model assumes that grains are completely spherical, with a radius $r_{\mathrm{g}}$, connected with concave bonds of radius $r_{\mathrm{b}}$.
The initial bond radius is set by the radius of contact when two grains come into contact.
For ice grains, this adhesion stage occurs due to the Van der Waals interaction between particles, leading to the expression of the bond radius \citep{Molaro2019}:
\begin{equation}
    r_{\mathrm{b}} \approx \left( \dfrac{\gamma r_{\mathrm{g}}^2}{ 10 \mu} \right)^{1/3}
    \label{eq:bondVDW}
\end{equation}
where $\gamma$ is surface tension of water ice and $\mu$ its shear modulus given in Table \ref{tab:properties}.

Following \cite{Swinkels1981, Molaro2019} with no densification, the radius of the imaginary circle between grains (see Figure \ref{fig:meta_geom}) is geometrically defined as
\begin{equation}
	r_{\mathrm{\rho}} = \dfrac{r_{\mathrm{b}}^2}{2(r_{\mathrm{g}} - r_{\mathrm{b}})}. 
    \label{eq:rho_radius}
\end{equation}
The $\hat{x}$ axis follows the direction from the grain center towards the bond center. 
The model considers the grain and bond separately; the transition between the two occurring at a distance $\hat{x}^{*}$ from the grain center:
\begin{equation}
	\hat{x}^{*} = r_{\mathrm{g}} \sin \theta^{*} 
\end{equation}
where $\theta^{*}$ is given by: 
\begin{equation}
	\theta^{*} = \arctan{\dfrac{r_{\mathrm{g}}}{r_{\mathrm{b}}+r_{\mathrm{\rho}}}}.
\end{equation}

The surface curvature is defined as the inverse of the radius of curvature.
The grain area, left of $\hat{x}^{*}$, is considered to have a constant curvature $K_{\mathrm{g}}$, given by
\begin{equation}
    K_{\mathrm{g}} = \frac{2}{r_{\mathrm{g}}}.
\end{equation}
The bond area, right of  $\hat{x}^{*}$, is assumed to have a constant curvature $K_{\mathrm{b}}$ calculated by the sum of curvatures along the two relevant axes ( $\hat{x}$ in red and $\hat{y}$ in blue on Figure \ref{fig:meta_geom})
\begin{equation}
    K_{\mathrm{b}} = \frac{1}{r_{\mathrm{b}}} - \frac{1}{r_{\mathrm{\rho}} }.
    \label{eq:curv_b}
\end{equation}
In reality, the transition between the grain and bond is smooth, with local changes in curvature near the interface. However, given the smaller bond radius compared to the grain radius, the surface affected by such changes is relatively small and thus not considered. Nonetheless, exploring this aspect in further studies could prove valuable, particularly when the bond radius approaches the size of the grains.

\subsection{Vapor Transport Diffusion}

\begin{figure}[htpb]
	\centering
	\includegraphics[width=0.8\textwidth]{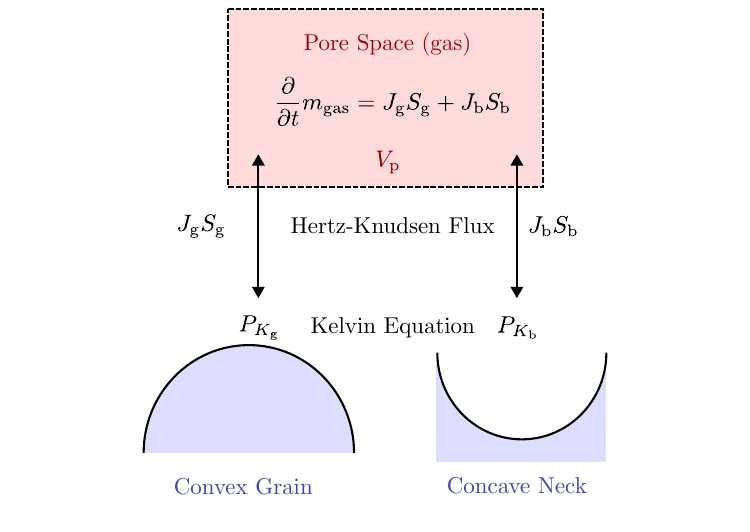}
 \caption{
    Schematic representation of diffusion by vapor transport. The grain and bond surfaces exchange matter with the surrounding gas in the pore space.
    }
    \label{fig:meta_model}
\end{figure}

The pore space is modeled as a reservoir  of volume $V_{\mathrm{p}}$ and an initial gas pressure $P_{\mathrm{gas}}(t=0)$ (see \autoref{fig:meta_model}). 
The relationship between the gas pressure $P_{\mathrm{gas}}$ and total mass of matter in the box, $m_{\mathrm{gas}}$, is given by the ideal gas law
\begin{equation}
	P_{\mathrm{gas}} = \frac{m_{\mathrm{gas}} RT}{V_{\mathrm{p}} M}.
	\label{eq:perfect_gas}
\end{equation}
where $M$ is the molar mass of water.
The gas mass variations in the pore space result from the flux of matter coming from the neighboring grain and bond and will continue to change until reaching a steady state.

One of the key benefits of such modelization is that no assumptions are made regarding the initial gas pressure. 
For example if the initial gas pressure is zero, both the grain and bond will sublimate to fill the pore space.
The model does not enforce a particular direction of transport, it is only when the gas pressure in the pore space approaches the saturation vapor pressure over a flat surface that matter will  naturally sublimate from the convex grain surface and condense on the concave bond surface.
%SF saturated vapor pressure => saturation vapor pressure

\subsubsection{Reaching the steady state}
Our model computes the gas mass variations over time using the sum of the Hertz–Knudsen flux ($J_{\mathrm{g}}$, $J_{\mathrm{b}}$) multiplied by their respective area of exchange ($S_{\mathrm{g}}$, $S_{\mathrm{b}}$)
\begin{equation}
	\frac{\partial }{\partial t} m_{\mathrm{gas}} = J_{\mathrm{g}}  S_{\mathrm{g}} + J_{\mathrm{b}}  S_{\mathrm{b}}.
	\label{eq:dmdt}
\end{equation}
The mass fluxes arriving in the gas phase are calculated using the Hertz–Knudsen equation \citep{Schrage1953, Jones2018}, from the difference between the gas vapor pressure $P_{\mathrm{gas}}(t)$ and the modified saturated vapor pressure due to curvature $P_{K_{\mathrm{j}}}$:
\begin{equation}
	J_{\mathrm{j}} = \alpha \left( P_{K_{\mathrm{j}}} - P_{\mathrm{gas}}(t) \right) \sqrt{\frac{M}{2 \pi R T}} 
    \label{eq:Hertz-Knudsen}
\end{equation}
where the subscript $j$ represent the grain or bond, $\alpha$ is the sticking coefficient,  $R$ the gas constant.
The modified saturated vapor pressure due to curvature $P_{K_{\mathrm{j}}}$ is  obtained from Kelvin equation \citep{Butt2003}
\begin{equation}
	P_{K_{\mathrm{j}}} \approx P_{\mathrm{s}}(T) \left( 1 + \frac{\gamma M}{R T \rho_{\mathrm{0}}} K_{\mathrm{j}} \right).
    \label{eq:Kelvin}
\end{equation}
where $\rho_{\mathrm{0}}$ is the bulk ice density  and $P_{\mathrm{s}}(T)$ is the saturated water vapor pressure over a flat surface, given by \citet{Feistel2007}.

Using Equations \ref{eq:Hertz-Knudsen} and \ref{eq:Kelvin} we can re-write  Equation \ref{eq:dmdt} as
\begin{align}
	\frac{\partial }{\partial t} m_{\mathrm{gas}} &=  \alpha P_{\mathrm{s}}(T) \left[ S_{\mathrm{b}} + S_{\mathrm{g}} + \frac{\gamma M}{R T \rho_{\mathrm{0}} }\left( S_{\mathrm{b}} K_{\mathrm{b}} + S_{\mathrm{g}} K_{\mathrm{g}} \right)  \right] \frac{2}{\pi \langle v \rangle} \\  \notag
										&- \alpha P_{\mathrm{gas}}(t) (S_{\mathrm{b}} + S_{\mathrm{g}}) \frac{2}{\pi \langle v \rangle}.
\end{align}
%vmean = sqrt{8RT/\pi Ml} # Wiki
where $\langle v \rangle = \sqrt{8 RT / M \pi}$ is the mean velocity of gas particles following a Boltzmann distribution.
Using Equation \ref{eq:perfect_gas}, this leads to a first-order differential equation for the gas mass
\begin{equation}
	\frac{\partial }{\partial t} m_{\mathrm{gas}} +  \frac{m_{\mathrm{gas}}}{\tau_{\mathrm{gas}}} = \frac{m_{\infty}}{\tau_{\mathrm{gas}}}
\end{equation}
with general solutions of the form 
\begin{equation}
	m_{\mathrm{gas}}(t) = C \,  e^{-t/\tau_{\mathrm{gas}}} +  m_{\infty}
\end{equation}
where $C$ is a constant to be determined.
From this formulation, appears a characteristic timescale for the gas mass variations $\tau_{\mathrm{gas}}$,  expressed as
\begin{equation}
	\tau_{\mathrm{gas}} = \dfrac{ 4 V_{\mathrm{p}}}{ \alpha (S_{\mathrm{g}}+S_{\mathrm{b}}) \langle v \rangle}
\end{equation}
and the gas mass at the steady state $m_{\infty}$ given by
\begin{equation}
	m_{\infty} =  \alpha  \tau_{\mathrm{gas}}   P_{\mathrm{s}}(T) \left[ S_{\mathrm{b}} + S_{\mathrm{g}} + \frac{\gamma M}{R T \rho_{\mathrm{0}} }\left( S_{\mathrm{b}} K_{\mathrm{b}} + S_{\mathrm{g}} K_{\mathrm{g}} \right)  \right] \frac{2}{ \pi \langle v \rangle}
    \label{eq:mgas_infty}
\end{equation}
By obtaining $m_\infty$ we get access to the gas pressure at the steady state as computed in Appendix \ref{sec:app_ssflux}, which will be used to compute the mass transfer in the steady regime.

%%%%%%%%%%
\subsection{Incrementing the Grain and Bond Masses}

Now that the gas pressure in the steady state has been obtained, the sublimation and condensation flux can be calculated using Hertz–Knudsen flux (Equation \ref{eq:Hertz-Knudsen}) and the vapor pressure Kelvin Equation  (\ref{eq:Kelvin}) for the grain or bond curvature.
To maintain numerical stability, it is necessary that the sintering timestep  is significantly smaller than the characteristic bond growth timescale: $\Delta t_{\mathrm{sint}} << \tau_{\mathrm{sint}}(r_{\mathrm{b}})$. This ensures that only a small fraction of grain and bond mass is altered, allowing us to assume constant grain geometry throughout a single iteration.
The updated grain and bond masses $m_{\mathrm{g}}$ and $m_{\mathrm{b}}$ are calculated using
\begin{equation}
	m_{\mathrm{j}}(t+\Delta t_{\mathrm{sint}}) = m_{\mathrm{j}}(t) - J_{\mathrm{j}}(t) S_{\mathrm{j}} \Delta t_{\mathrm{sint}}
	\label{eq:update_mass}
\end{equation}
where $J_{\mathrm{j}}$ is the Hertz–Knudsen flux in the steady state, with a negative sign because it represents mass entering the pore space and $S_{\mathrm{j}}$ is the surface area in contact with the interface.
The analytical expression of the surfaces areas in contact with the phase exchange are determined using geometrical considerations (see Appendix \ref{sec:app_twobonds}), knowing that due to symmetry only a quarter of the total grain/model system needs to be represented. 

\subsection{Updating the geometry}

Once new masses have been calculated for the grain and bond, the corresponding volumes $V_{\mathrm{g}}, V_{\mathrm{b}}$ are numerically obtained by multiplying these masses with the bulk density of ice
\begin{equation}
  V_{\mathrm{j}}(t+\Delta t_{\mathrm{sint}})  = \dfrac{m_{\mathrm{j}}(t+\Delta t_{\mathrm{sint}})}{\rho_{\mathrm{0}}}
\end{equation}
To retrieve the resulting new grain and bond radii $r_{\mathrm{g}}$, $r_{\mathrm{b}}$, we need to obtain the relationship between the radii and volumes.
Using geometrical considerations, we can derive the analytical relationship between the grain and bond radii and their respective volumes $v_{\mathrm{g}}$, $v_{\mathrm{b}}$ (see Appendix \ref{sec:app_twobonds}). 

The issue is that  there is no direct analytical equation that gives the grain or bond radii $r_{\mathrm{g}}, r_{\mathrm{b}}$ as  functions of their volumes.
To address this, from the analytical expression of the grain volume  (\autoref{eq:app_grain_vol}) and bond volume (\autoref{eq:app_bond_vol}) we use the Newton-Raphson method to find the updated values of $r_{\mathrm{g}}$ and $r_{\mathrm{b}}$.
This method finds the roots of the difference between the numerically computed volumes $V$ and the analytically obtained volumes $v$:
\begin{align}
	&r_{\mathrm{g}}(t+\Delta t_{\mathrm{sint}}) = \min_{r_{\mathrm{g}}} \left| V_{\mathrm{g}}(t+\Delta t_{\mathrm{sint}}) - v_{\mathrm{g}}(r_{\mathrm{g}}, r_{\mathrm{b}}(t)) \right| \\
 	&r_{\mathrm{b}}(t+\Delta t_{\mathrm{sint}}) = \min_{r_{\mathrm{b}}} \left| V_{\mathrm{b}}(t+\Delta t_{\mathrm{sint}}) - v_{\mathrm{b}}(r_{\mathrm{g}}(t), r_{\mathrm{b}}) \right| .
\end{align}
Once the Newton-Raphson algorithm gives the updated values of the grain and bond radii, the other geometric parameters $r_{\mathrm{\rho}} $, $\theta^{*}$, $\hat{x}^{*}$, $K_{\mathrm{g}}$ and $K_{\mathrm{b}}$ are updated accordingly.
Upon finding the updated geometry, the algorithm is ready to determine a new steady state and repeat the process iteratively.

\subsection{Comparison with Experimental Data}

To confront our analytical model against real-world sintering scenarios, we compare it to a well-known experiment on ice sintering conducted by \citet{Hobbs1964}. This comprehensive study involved spherical particles ranging from $50$ to $\SI{700}{\micro \meter}$ in diameter, exposed to temperatures between $\SI{-3}{\degree C}$ and $\SI{-20}{\degree C}$.
After approximately one hour, bond growth was observed, regulated by the vapor supply originating from nearby regions  \citep{Blackford2007}.
Based on their findings, the authors concluded that vapor transport serves as the dominant mechanism in the sintering process. 

\begin{figure}[htpb]
	\centering
	\includegraphics[width=0.6\textwidth]{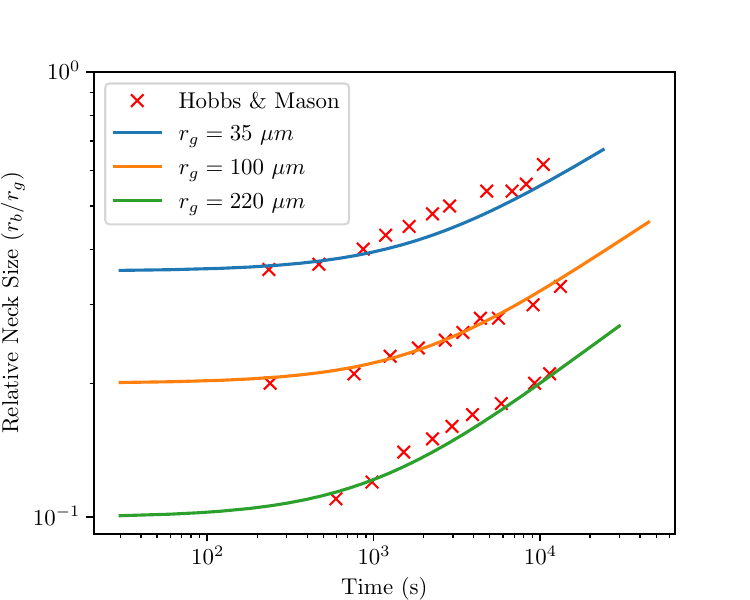}
 \caption{ Evolution of the relative bond radius $r_{\mathrm{b}}/r_{\mathrm{g}}$ in \citet{Hobbs1964} experiments (crosses) for different grain sizes at a constant temperature $\SI{-3}{\degree C}$. 
 With a single value of the sticking coefficient $\alpha = 0.03$, the model (lines) can accurately compute the relative bond radius evolution for the three experiments.
    }
    \label{fig:valid_meta}
\end{figure}

The evolution of the relative bond radius from the $\SI{-3}{\degree C}$ experiments, conducted for grain radii of $\SI{35}{\micro\meter}$, $\SI{100}{\micro\meter}$, and $\SI{200}{\micro\meter}$, is illustrated in Figure \ref{fig:valid_meta} (marked with crosses).
To compare our model with the experiments, we need to assign a value to the sticking coefficient $\alpha$, which is poorly known but we take the advantage of these experimental data to estimate it.  By setting our sticking coefficient value to $\alpha = 0.03$, we are able to effectively fit all the experiments, regardless of grain size.

The sticking coefficient value used in our model is notably low compared to values reported in the literature, such as 0.75 at 140 K \citep{Haynes1992}. According to the extrapolation of \citet{Haynes1992}, the sticking coefficient would be expected to be around 0.5 at the temperature of the experiment of $\SI{-3}{\degree C}$ used in Figure \ref{fig:valid_meta}. Despite also fitting the data at $\SI{-10}{\degree C}$, $\SI{-15}{\degree C}$, and $\SI{-20}{\degree C}$, the sticking coefficient remains very low  without a clear trend as the temperature decreases. Therefore, in our model, the sticking coefficient functions more as a fitting parameter rather than reflecting a physical property.

The major difference is that \citet{Hobbs1964} conducted experiments at atmospheric pressure ($\sim \SI{1}{bar}$), while our model assumes that the pores are only filled with water vapor sublimated from the grains. 
In air, water vapor diffusion would be significantly limited due to a much smaller mean free path compared to  near vacuum conditions, where water molecules can move almost directly from the grains to the bonds. So, at atmospheric pressure, diffusion creates a vapor pressure gradient between the sublimating surface of the grain and the condensing surface of the bond. This gradient results in a gas pressure $P_{\mathrm{gas}}$  that is different near the grain and near the bond, unlike in a near vacuum conditions.
As a result, the pressure difference $\Delta P = P_{K_\mathrm{j}} - P_{\mathrm{gas}}$ in the Hertz-Knudsen Equation \ref{eq:Hertz-Knudsen} is significantly smaller in the confined air layer close to each surface. Therefore, the sticking coefficient we fitted is approximately 17 times smaller than expected, likely because the pressure difference $\Delta P$ is about 17 times smaller at atmospheric pressure than in near vacuum.

However, the ability to use a single value of the sticking coefficient to fit the three experiments conducted under different grain size conditions  is indicative of a consistent physical description, demonstrating robustness in the model's approach across varying grain sizes. This agreement also validates our approach.

\subsection{Varying Number of Bonds per Grain}

The previously defined geometry assumed that each grain is connected to its neighbors by two bonds, while in real cases, the number of bonds per grain may vary.
To estimate this influence, we adapted our model to three scenarios: a grain connected by a single bond, by two bonds as initially presented, and by six bonds (two in each direction).
These modifications change the geometry of the grain-bond system, with detailed surface and volume derivations provided in Appendix \ref{sec:app_geom}.

\begin{figure}[htpb]
	\centering
    \hspace*{-1.1cm}
	\includegraphics[width=1.0\textwidth]{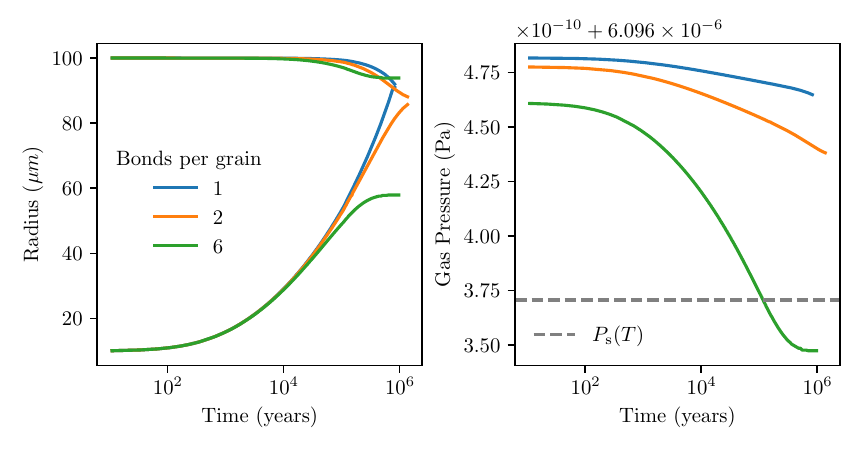}
    \caption{(\textit{Left}) Evolution of grain radius (top) and bond radius (bottom) over time at $T = \SI{150}{K}$ for different bond-per-grain scenarios (1, 2, and 6).
(\textit{Right}) Evolution of gas pressure $P_{\mathrm{gas}}$ over time for different bond-per-grain scenarios (1, 2, and 6). The gray dashed line represents the saturated vapor pressure over a flat surface at $T = 150 K$ .}
    \label{fig:bondpergrain}
\end{figure}

Using a reference scenario at a fixed temperature of $T = \SI{150}{K}$ and a grain size of $r_{\mathrm{g}} = \SI{100}{\micro m}$, we plotted the different evolutions of the grain and bond radii over time for varying bonds-per-grain values in Figure \ref{fig:bondpergrain} (\textit{Left}).
Firstly, the results show that up to about half of the bond's final size, its evolution does not depend on the number of bonds per grain. However, in the late stage of sintering, differences emerge.
For the two bonds per grain case, after significant time, the surface of the bond-grain system becomes almost completely flat with the bond radius nearing that of the grain.
In the six bonds per grain scenario, the late stage of sintering is limited by competition among bonds for condensation, resulting in an equilibrium where the bond radius is smaller than the final grain radius.
In the single bond per grain scenario, the bond growth does not follow an asymptote. Our explanation is that while one hemisphere of the grain connected to the bond will eventually flatten, the other hemisphere, not connected to a bond, remains spherical. Thus, sublimation of the  unconnected hemisphere continues to provide matter for condensation over the flat bond.
Note that, simulations were stopped when the bond radius reached values too close to the grain radius, as this induces computational errors.

The simulations also allow us to track the evolution of gas pressure over time in the pore space for different bonds per grain scenarios, as shown in Figure \ref{fig:bondpergrain} (\textit{Right}).
Firstly, we observe that the equilibrium gas pressure is always different from the saturated vapor pressure over a flat surface. In most scenarios, the gas pressure is higher than the saturated vapor pressure, with fewer bonds per grain leading to more convex surfaces and consequently higher gas pressure. Interestingly, in the case of six bonds per grain, when the bonds reach a critical radius (after around 100,000 years in Figure \ref{fig:bondpergrain}), the gas vapor pressure drops below the saturated vapor pressure over a flat surface. In this case, the system becomes dominated by concave surfaces formed by the bonds, resulting in lower gas pressure.

While these late-stage evolution scenarios provide new results on the sintering behavior under different grain environment, they should be considered with caution, as other diffusion mechanisms could become predominant at this stage (further discussed in Section \ref{sec:porosity}), and vapor transport sintering may become irrelevant after a certain bond threshold \citep{Molaro2019}.  \\

\subsection{Sintering Timescales}

\begin{figure}[htpb]
	\centering
	\includegraphics{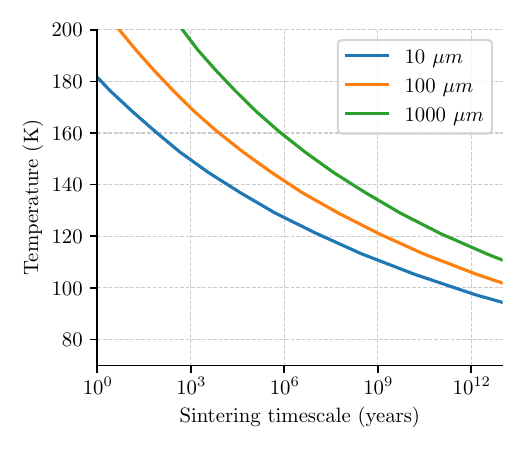}
 \caption{Sintering timescales for different temperatures and initial grain radii, computed based on simulations with two bonds per grain. The timescale is determined when the bond radius reaches half of the grain radius.}
    \label{fig:sinter_timescale}
\end{figure}

The equivalent behavior of the different bond-per-grain scenarios in the initial to mid stages of sintering allows us to define a unique characteristic sintering timescale irrespective of the number of bonds per grain.
We choose to define the sintering timescale as the time it takes for the bond radius to reach 50\% of the grain radius, in a two bond-per-grain case:
\begin{equation}
r_{\mathrm{b}}(\tau_{\mathrm{sint_{50}}}) = \frac{1}{2} r_{\mathrm{g}}(\tau_{\mathrm{sint_{50}}}).
\end{equation}
By running the sintering model, we can retrieve the sintering timescales for different grain sizes and temperatures, as shown in Figure \ref{fig:sinter_timescale}. Note that the results depend on the arbitrary definition of the sintering timescale. If we defined this timescale as the time it takes for the bonds to reach 90\% of the grain radius or or the time it takes for the bonds to double their initial size, the timescales would differ significantly.
These definitions provide first-order estimates at a fixed temperature, but more detailed and accurate results require simulating sintering coupled with a thermal solver.

Another way to analytically estimate a characteristic timescale for sintering $\tau_{\mathrm{sint_V}}$, would be to defined it as the time for condensation to significantly change the bond volumes:
\begin{equation}
	\tau_{\mathrm{sint_V}}(r_{\mathrm{b}})	= V_{\mathrm{b}}  \left| \frac{\partial t}{\partial V_{\mathrm{b}}} \right|.
\end{equation}
The variations of the bond volume due to condensation can then be related to the condensation flux on the bond surface
\begin{equation}
	\frac{\partial V_{\mathrm{b}}}{\partial t} = \dfrac{1}{\rho_{\mathrm{0}}} \frac{\partial m_{\mathrm{b}}}{\partial t}   = \frac{J_{\mathrm{b}} S_{\mathrm{b}}}{\rho_{\mathrm{0}}}.
\end{equation}
Developing the Hertz–Knudsen equation once the gas has reached steady state (see Appendix \ref{sec:app_ssflux}) leads to an analytical expression of the characteristic bond growth timescale
\begin{equation}
		\tau_{\mathrm{sint_V}}(r_\mathrm{b}) = \sqrt{2 \pi}    \dfrac{  \rho_{\mathrm{0}}^2 V_{\mathrm{b}}  S }{ \alpha \gamma P_{\mathrm{s}} S_{\mathrm{b}} S_\mathrm{g} (K_\mathrm{b} - K_\mathrm{g})}  \left( \dfrac{R T}{M} \right)^{3/2}.
\end{equation}

However, the explicit dependence on the bond radius underscores that this characteristic time  is not constant throughout the sintering process. Rather, it provides an indication of the time required to observe noticeable growth of the bond volume.

While, two analytical timescales ($\tau_{\mathrm{sint_V}}(r_\mathrm{b}), \tau_{\mathrm{gas}}$) appear in our description of the sintering process, they characterize different physics and vary on vastly different scales.
For instance, at a temperature of $\SI{140}{K}$, with a grain radius of $r_{\mathrm{g}} = \SI{100}{\micro\meter}$, a bond radius of $r_{\mathrm{b}} = \SI{10}{\micro\meter}$, and a sticking coefficient of $\alpha = 0.03$, the characteristic bond growth timescale is $\tau_{\mathrm{sint_V}}(r_\mathrm{b}) \sim \SI{3600}{years}$, while the characteristic timescale of gas mass variations is $\tau_{\mathrm{gas}} \sim \SI{e-11}{s}$. 
Hence while the steady state is reached almost instantly, the sintering timescale takes much longer to occur.

\section{Multiphysics Coupling with LunaIcy}

%The sintering process is highly dependant in temperature.  
%So, following suggestion by \citep{Molaro2019}, to accurately simulate the effect of sintering, the precise temperature at any given time needs to be known.
%For such reasons, 

\begin{figure}[htpb]
	\centering
\includegraphics[width=0.9\textwidth]{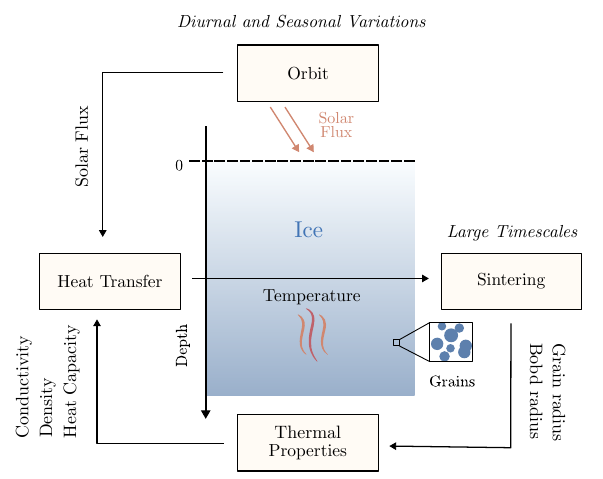}
 \caption{Block diagram of the proposed multi-physics simulation model LunaIcy.}
\label{fig:diag_model}
\end{figure}

To accurately simulate the temperature-dependent sintering process, the sintering model is coupled with the thermal solver MultiHeaTS \citep{Mergny2024h} creating our multiphysics simulations model, LunaIcy (see Figure \ref{fig:diag_model}).
The model consist of a uni-dimensional bloc of ice made of grains, with a certain porosity and thermal properties for each depth. 
In this numerical model, we use an irregular spatial grid consisting of $n_x$ points, which we iterated for a total of $n_t$ iterations. 
Here the vertical axis $x$ is the same as the $\hat{x}$ axis  of the sintering model.
However, the variables have different names to clearly indicate the important distinction: $x$  represents the depth, while $\hat{x}$ represents the distance to the grain center in the sintering model.

%timescales

The spatial and temporal parameters are discretized as follows:
\begin{equation}
\begin{cases}
    x \rightarrow x_n &= x_{n-1} + \Delta x_n \\
    t \rightarrow t^i &= t^{i-1} + \Delta t^i .\\
    
\end{cases}
\end{equation}
Here, $n$ is an integer such that $n \in \{0, \: \dotsc \: , n_x-1 \} $, representing the $n$th element in the spatial dimension, and $i$ is an integer such that $i \in \{0, \: \dotsc \: , n_t-1 \} $, representing the $i$th element in the time dimension. 
It is worth noting that both the spatial $\Delta x_n$ and temporal $\Delta t^i$ increments may not be constant.
This flexibility is advantageous as it allows for the use of irregular spatial grids and optimization of the timestep to be more computationally efficient.

By calculating the orbit of the target body, we determine the solar flux heating the ice surface. Then, our thermal solver MultIHeaTS, computes the heat transfer throughout the material's depth for each timestep. 
At each depth, the sintering module assesses the microstructure changes corresponding to the current temperature,  and then updates the ice's thermal properties accordingly.
Since the pore spaces do not communicate, the grain/bond system forms a closed system. Therefore, each layer of LunaIcy contains a grain/bond system that computes sintering independently based on its layer properties.
This iterative process is repeated throughout the simulation. A summary of the thermal modules is provided here, with a detailed description available in the joint paper \citep{Mergny2024h}.

\subsection{Heat Transfer}

\paragraph{Temperature}
MultiHeaTS proposes to solve numerically the heat equation where all thermal properties can vary continuously over space and time, and with non-constant time and space increments. The derivation of the fully implicit scheme and additional details can be found in \cite{Mergny2024h}.
In one dimension, the heat equation for conduction transfer can be expressed as:
\begin{equation}
    \rho(x, t) c_{\mathrm{p}}(x, t) \dfrac{\partial T(x, t)}{\partial t} =  \dfrac{\partial}{\partial x} \left( k(x, t) \dfrac{\partial}{\partial x} T(x, t) \right)
    +  Q(x, t)
    \label{eq:heat-equation}
\end{equation}
where $\rho$ is the density, $c_{\mathrm{p}}$ is heat capacity, $k$ is thermal conductivity, and $Q$ denotes an optional source or sink term.
The upper boundary condition (i.e., the flux leaving the surface) is determined by the energy equilibrium between the solar flux and the gray body emission from the surface \citep{Spencer1989}:
\begin{equation}
    \forall t, k(0,t) \left. \dfrac{\partial T(x, t)}{\partial x}\right|_{x=0} =  - F_{\mathrm{solar}}(t) + \epsilon \cdot \sigma_{\mathrm{SB}} \cdot T(0, t)^4
    \label{eq:energy-eq}
\end{equation}
where $\epsilon$ is the thermal emissivity and $\sigma_{\mathrm{SB}}$ the Stefan-Boltzmann constant. 

The expression of the solar flux is based on \cite{Spencer1989} as a function of latitude $\lambda$ and longitude $\psi$:
\begin{equation}
	F_{\mathrm{solar}}(t, \lambda, \psi) = 
			\left(1 - A(\lambda, \psi) \right) \dfrac{G_{\mathrm{SC}}}{d(t)^2} \cos(\theta_i(t,\lambda, \psi))
 \label{eq:slr_flux_model_accurate}
\end{equation}
where $A$ is the surface albedo, $G_{\mathrm{SC}}$ the solar constant, $\theta_i$ the solar incidence angle and $d$ the distance to the Sun in AU, estimated using harmonic functions as described in \cite{Mergny2024h}.
%Thanks to its stability and computational efficiency, MultIHeaTS is particularly advantageous for simulating processes that occur on large timescales such as sintering on icy moons.

By default, and hereafter, the thermal solver is set without considering eclipses, which is strictly valid for the anti-Jovian hemispheres of the Galilean moons. 
However, it is possible to extend these results to any longitude by accounting for the eclipse effect at first order using an equivalent albedo \citep{Mergny2024h} (Equation 40 of the cited article, with less than 0.5\% error on the temperature distribution).

\paragraph{Porosity}
\label{sec:porosity}
While sintering could lead in some cases to a change of density, it remains unclear when such densification occurs.
The sintering mechanism mostly studied under planetary conditions is diffusion by vapor transport, a non-densifying mechanism \citep{Swinkels1981, Blackford2007, Molaro2019}. 
This process redistributes mass from grains to bond areas but does not reduce pore space. 
While it strengthens the mechanical properties of the ice by cementing grains together, it does not change density.

If densification does occur, it would likely involve a different diffusion mechanism like grain boundary diffusion \citep{Kaempfer2007}, particle-induced compaction 
\citep{Raut2008, Schaible2017} or what is referred as "late-stage" sintering by \citet{Molaro2019},  possibly operating over longer timescales than vapor diffusion.
Although work by \cite{Eluszkiewicz1991} proposed a model for ice densification through sintering, their expression is derived from “an unpublished internal report by Ashby 1988” which existence could not be verified \citep{Molaro2019}.
On Earth, some forms of sintering can typically increases density due to the recrystallization of liquid water in the pore space. 
However, in dry sintering, it is not clear if a densification happens.
Therefore, while various other compaction mechanisms can intervene (e.g. sputtering or  late stage sintering) we currently lack the ability to quantify their link to densification.

For these reasons, to model the density profile, we refer to our previous work \citep{Mergny2024c}, that specifically focuses on ice compaction driven by overburden pressure. 
Given that the thermal skin depth on Europa is estimated to be a few centimeters \citep{Mergny2024h}, whereas the compaction length scale is on the order of hundreds of meters \citep{Mergny2024c}, it is reasonable to assume that the porosity remains constant at the depths relevant to our thermal analysis.
This lead to the simplified expression for the density of porous ice
\begin{equation}
    \rho(\phi) = \rho_{\mathrm{0}} (1- \phi).
\end{equation}

\paragraph{Conductivity}

In contrast, sintering drastically changes the thermal conductivity of ice.
As bond growth occurs, the surface area of solid to solid increases leading to an increase in conductivity. 

Expressions found in the literature place significant importance in the effect of the bond radius on the effective conductivity of the material \citep{Adams2001, Gundlach2018, Lehning2002b}.
For these reasons, following \citep{Mergny2024h}, we use the expression of the porous conductivity:
\begin{equation}
    k(\phi, r_{\mathrm{b}}, r_{\mathrm{g}}) = k_{\mathrm{0}}(T) \left(1-\phi \right) \frac{r_{\mathrm{b}}}{r_{\mathrm{g}}}.
    \label{eq:cond_porous}
\end{equation}
where the bulk crystalline ice conductivity for a given temperature is computed using the expression $k_{\mathrm{0}} = 567/ T$ \citep{Klinger1981}.
Updating the temperature-dependent conductivity for each iteration would slow down computation significantly, so we maintain a constant bulk ice conductivity based on the surface equilibrium temperature.
While temperature variations on Europa do not have a significant impact on conductivity compared to porosity and grain contact changes, it would be valuable to include in other applications.

Europa's icy surface is expected to be a mixture of crystalline and amorphous phases, at least in the first millimeter depth 
\citep{Hansen2004, Ligier2016, Berdis2020}.
However, consistency among different lab measurements on amorphous ice conductivity has not been reliable \citep{Gudipati2013}, likely due to the wide range of porosity and structural diversity in these experiments \citep{Prialnik2022}.
For porosity-independent modeling, it would be helpful to experimentally measure the thermal conductivity of ice before and after amorphization or thermal relaxation.
In the absence of such data, we consider only crystalline ice and will address model improvements in future versions of LunaIcy.

As the bond and grain radii evolve due to sintering throughout the simulation, the conductivity is adjusted according to Equation \ref{eq:cond_porous} to reflect these microstructural changes. This updated conductivity is then accounted in the thermal module, closing the two-way coupling between sintering and heat transfer.

\subsection{Initialization and parameters}

The LunaIcy model is run over a range of parameters to simulate the evolution of Europa's icy surface microstructure under various conditions.
While temperature variations occur within the diurnal period, sintering progresses over large timescales under Europa's conditions, leading to a numerical challenge that can be handled with LunaIcy.

To capture temperature variations, the simulations require a precise timestep $\Delta t = p_{\mathrm{E}} / n_{\mathrm{spd}}$, where $p_{\mathrm{E}} = \SI{3.55}{days}$ is Europa's orbital period and the number of steps per day is set to $n_{\mathrm{spd}} = 30$, balancing accurate computations with efficient computation time.
Over the one million year timescale, this results in a total number of iterations $n_t > \SI{3e9}{}$.
To address this, the unconditional stability property of the implicit scheme used in MultIHeaTS, coupled with its capability to use an irregular grid, decrease the computation time of each iteration.
In these simulations, the spatial increments $\Delta x_n$ are computed on an irregular grid given by the relation
\begin{equation}
   \forall n \in \{0, \: \dotsc \: , n_x-1 \} , \,  x_n = \left( \frac{n}{n_x -1} \right)^{5} L
\end{equation}
where $L$ is the maximum depth computed in the model, and the increase in spatial step with depth was chosen for efficiency reasons, as detailed in the joint article \citep{Mergny2024h}. 
Given that sintering progresses much slower than temperature variations, the grain and bond radii are only updated after significant amount of time has elapsed, set to $\Delta t_{\mathrm{sint}} = 0.01 \times \tau_{\mathrm{sint_V}}(r_{\mathrm{b}})$.
This enables calling the heat transfer module more frequently than the module responsible for updating thermal properties, as shown in Figure \ref{fig:diag_model}. 
Thanks to such optimization, particularly due to the thermal solver, the computation time of a one million year simulation was reduced to around 4 days on an 6 cores Intel i7-10750H CPU at 3.6 GHz. 

Given the large number of iterations $n_t$, saving the ice state frequently would take significant storage space. 
To solve this problem, the properties are saved periodically, every $p_{\mathrm{save}} = 10.1 \times p_\mathrm{J}$, where $p_\mathrm{J} = \SI{11.86}{years}$ is Jupiter's current orbital period. 
This ensures that we capture all diurnal and orbital variations while maintaining a manageable file size, typically around 1 GB per simulation.

The determination of initial ice structure is difficult due to lack of information on Europa. The origin of the surface porosity is not clear; it could be the product of a plume deposit or the result of crushed ice by space weathering. Here, we propose to assume a relatively loose granular material, with parameters in agreement with observed surface temperature from infrared measurement \citep{Rathbun2010, Trumbo2018} and grain size in agreement with near-infrared spectroscopic measurement \citep{Hansen2004, CruzMermy2023}.

\begin{table}[htbp]
    \centering
    \begin{tabular}{lr}
    \toprule
    Parameter & Value   \\ 
    \midrule
    Final time $t_{\mathrm{f}}$ & \SI{1}{My}\\
    Timestep $\Delta t$ & \SI{170}{min}\\
    Steps per day $n_{\mathrm{spd}}$ &   30   \\
    Grid points $n_x$ & 30  \\
    Max depth $L$   & 50 m    \\
    Latitude $\lambda$ & 0 \\
    Longitude $\psi$ & 0 \\
    Emissivity $\epsilon$ &  0.9  \\
    Albedo $A$ & 0.4 (0.4 - 0.8)   \\
    Porosity $\phi$ & 0.8 (0.2 - 0.9)  \\
    Sticking coefficient $\alpha$ &  0.03  \\
    Initial grain radius $r_{\mathrm{g}}(x, 0)$ & \SI{100}{\micro\meter} \\
    \bottomrule
    \end{tabular}
\caption{Physical and numerical parameters used for our simulations. The reference warm scenario is for an albedo of 0.4 and porosity of 0.8.}
\label{tab:parameters}
\end{table}

\section{Results and Discussion}

\subsection{Results}

\begin{figure}[htpb]
	\centering
    \hspace*{-35pt}
	\includegraphics{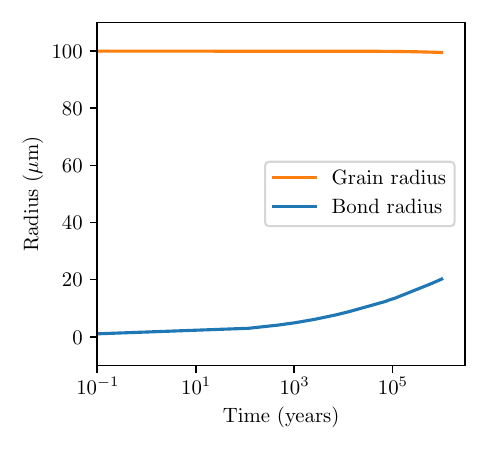}
     \caption{Top layer grain and bond radii evolution over one million years in a warm simulation scenario with a porosity of $\phi = 0.8$ and an albedo of $A=0.4$, resulting in a mean temperature of $\SI{114}{K}$ and a maximum daily temperature of $\SI{141}{K}$. The bond radius has increased by 59 times its initial value, leading to significant changes in the thermal, mechanical and optical properties of the ice. }
    \label{fig:meta_evo}
\end{figure}

We first present the results of a simulation under reference warm conditions for a typical grain radius of $r_{\mathrm{g}}(t=0) = \SI{100}{\micro\meter}$, a porosity of $\phi = 0.8$, an albedo of $A=0.4$, and the solar flux set at the equator, $\lambda = 0$.
Using the complete set of parameters given in Table \ref{tab:parameters}, we obtain a mean surface temperature of $\SI{114}{K}$ and a maximum daily temperature of $\SI{141}{K}$,  close to the highest values obtained from Galileo photopolarimeter–radiometer \citep{Rathbun2010}. 
The evolution of the top layer grain $r_\mathrm{g}(x=0, t),$ and bond  $r_\mathrm{b}(x=0, t)$ radii throughout the simulation is shown in Figure \ref{fig:meta_evo}. 
We can observe that significant sintering occurs under these warm conditions, with the top layer bond radius increasing from $\SI{0.39}{\micro \meter}$ to $\SI{23.05}{\micro \meter}$ over one million years. In contrast, the grain radius decreases from $\SI{100.00}{\micro \meter}$ to $\SI{99.96}{\micro \meter}$ during the same period.
This difference can be explained by the fact that although the matter condensing on the bonds originates from the grains, the initial volume of the grains is five orders of magnitude larger than that of the bonds. Thus, even minor observable changes in the grain radius result in significant increases in the bond radius.

These results highlights the necessity of the thermal coupling; without it, relying on constant temperatures would lead to inaccurate sintering estimates. 
For instance,  with bond size  $r_\mathrm{b} = \SI{10}{\micro \meter}$, at the max daily temperature of $\SI{141}{K}$, the characteristic sintering time is $\tau_{\mathrm{sint_V}} \sim \SI{2700}{years}$ while at the mean temperature of $\SI{114}{K}$,  $\tau_{\mathrm{sint_V}} \sim \SI{51}{My}$.

\begin{figure}[htpb]
    \centering
    \hspace*{-50pt}
	\includegraphics{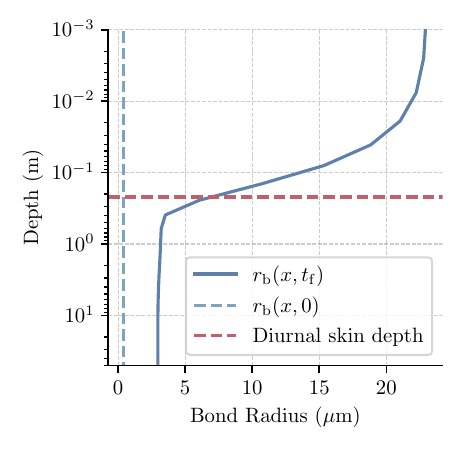}
     \caption{Bond radius versus depth after one million years of sintering with an albedo of $A = 0.4$ and a porosity of $\phi = 0.8$. The initial bond radius $r_\mathrm{b}(x, 0) \approx \SI{0.4}{\micro\meter}$, is indicated by blue dashed lines for reference. Significant sintering is observed up to the diurnal thermal skin depth.}
    \label{fig:bond_profile}
\end{figure}

However, sintering extends beyond the top layer, and the numerical simulations enable us to investigate its effects throughout the ice depth. 
In Figure \ref{fig:bond_profile} is plotted the bond radius profile versus depth after one million years, revealing significant sintering at depths shallower than the diurnal thermal skin depth (around $\SI{25}{cm}$ in this case). 
While the simulations started with an homogeneous structure, due to the damping of the thermal wave, sintering creates heterogeneity in the ice microstructure.
Such changes of the microstructure influence the macroscopic properties like the thermal conductivity as described by Equation \ref{eq:cond_porous}.

These results are of primordial importance when modeling the thermal properties of Europa for remote sensing; in warm regions, sintering is expected to occur and lead to an heterogeneous ice profile up to the thermal skin depth. As discussed in \cite{Mergny2024h}, this bilayer case could be observed by thermal infra-red observation.
This highlights the necessity of using multilayered thermal solvers, such as MultIHeaTS \citep{Mergny2024h} or the Planetary Code Collection \citep{Schorghofer2022} to derive the thermal properties of the surface.\\

The strength of this numerical approach is its ability to explore the evolution of an icy surface under different sets of parameters.
Europa's past and present surface properties are not well constrained, so this parameter exploration allow us to study the surface under different conditions.
In this study, we first conducted 25 simulations, varying porosities from 0.2 to 0.9 and albedo from 0.4 to 0.8, typical for Europa \citep{McEwen1986, Belgacem2020}.

\begin{figure}[htpb]
	\centering
	\includegraphics{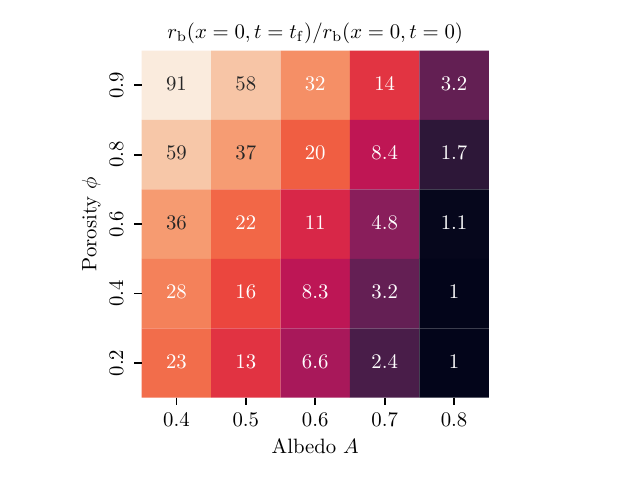}
     \caption{Sintering efficiency heatmap for the top layer as a function of porosity and albedo. Colors indicate the ratio of the top layer bond radius after one million years $r_\mathrm{b}(t=t_\mathrm{f})$ to the initial bond radius $r_\mathrm{b}(t=0)$. In the majority of scenarios, the bond radius more than doubles after one million years, except for very high albedo.}
    \label{fig:alb_poro}
\end{figure}

Figure \ref{fig:alb_poro} shows the sintering heatmap of the top layer (i.e., at the surface, depth $x=0$) for  this set of simulations; for each pair of the porosity and albedo parameters, is represented the ratio of the final bond radius to the initial one $r_{\mathrm{b}}(x=0, t_f)/r_{\mathrm{b}}(x=0, t=0)$. 
As expected, lowering the albedo or increasing the porosity, which yields a less conductive material, increases the surface temperature thereby enhancing sintering.
Only situations of very high albedo $A > 0.8$, do not see any effect of sintering during the one million year period.
For the warmest regions, the top layer bond radius grows by almost two orders of magnitude larger than its initial size, resulting in the same increase factor for the ice thermal conductivity (Equation \ref{eq:cond_porous}).

In most conditions represented here, except for very high albedo, the bond radius at least double its size after one million years. 
A consistent trend suggests that regions reaching a maximum daily temperature exceeding $\SI{115}{K}$ will experience noticeable sintering over this period. 
While the mean temperature of these surfaces is too cold for effective sintering, the probability distribution of temperature \citep{Mergny2024h} reveals that even when less than $5 \%$ of the time is spent above $\SI{115}{K}$, noticeable sintering can  occur over one million years.

The effect of longitude on the anti-jovian hemisphere is trivial and null given that it does not alter the solar flux. 
Latitude, however, has a straightforward effect: there is simply less solar flux on average but also at noon. Since the obliquity is near zero, the duration of the day is homogeneous over the moon, whatever the latitude. The influence of the latitude $\lambda$ can be approximated at first order by introducing an equivalent albedo $A_{\lambda}$ defined as:
\begin{equation}
    A_{\lambda} =  1 + (A-1)\cos(\lambda).
    \label{eq:latitude_albedo_relationship}
\end{equation}
For instance, the reference albedo used previously $A = 0.4$ would change to $A_{\lambda}=0.58$ at latitude $\lambda = \pm \SI{45}{\degree}$ and $A_{\lambda}=0.79$ at latitude $\lambda = \pm \SI{70}{\degree}$.
Thus, using the results from Figure \ref{fig:alb_poro}, limited sintering is expected for latitude poleward $\SI{70}{\degree}$ on Europa.
As a conclusion, the albedo map of Europa is expected to control the sintering of the ice, with an effective reduction toward the poles due to latitude effects.

\begin{figure}[htpb]
	\centering
	\includegraphics{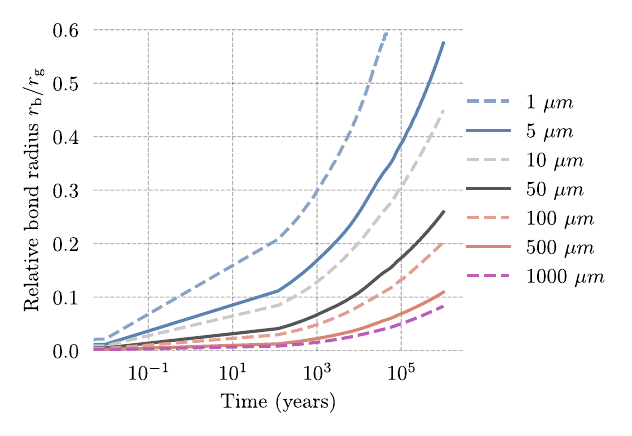}
     \caption{Evolution of the top layer relative bond radius $r_{\mathrm{b}} / r_{\mathrm{g}} $ for different initial grain radius $r_{\mathrm{g}}(t=0)$. Smaller grains show more efficient sintering due to their higher surface curvature. In situations of intense sintering, numerical instabilities start to appear after reaching a critical threshold $r_{\mathrm{b}} > 0.6 r_{\mathrm{g}} $.
     }
    \label{fig:grain_dep}
\end{figure}

A set of simulations was conducted to investigate the effect of grain sizes on the efficiency of sintering. The albedo and porosity were fixed to their values in the reference warm scenario outlined in Table \ref{tab:parameters}. 
Seven simulations were run with varying grain radii ranging from $\SI{1}{\micro \meter}$ to $  \SI{1}{mm}$. 
In Figure \ref{fig:grain_dep}, we present the evolution of the relative bond radius $r_{\mathrm{b}} / r_{\mathrm{g}} $ for the different grain sizes, focusing on the top layer, at $x=0$.
As anticipated, the sintering timescale shows a significant dependence on the grain size, with smaller grains undergoing faster sintering due to their higher surface curvature \citep{Blackford2007, Molaro2019}. 
While the exact grain size of Europa's surface remains unknown, thanks to such parameter exploration, we can predict the evolution of the various possible surface configurations.

The time of our simulations is shorter than the estimated ice crust age of approximately $\SI{30}{My}$ \citep{Pappalardo1998}, suggesting that, in present times, warm regions on Europa should have experienced significant sintering.
These numerical results are consistent with polarimetric comparisons of Europa with pure ice particles \citep{Poch2018}, suggesting that Europa is possibly covered by relatively coarse and sintered grains.
This is particularly noteworthy, as if these areas are made of an interconnected grain structures due to sintering, it drastically changes the thermal, optical and mechanical properties of the surface.
Incorporating these effects would be pertinent in designing a lander for Europa's surface.

\subsection{Discussion}

\subsubsection{Irregular grain shapes}
The current model simplifies the sintering system to two curvatures: the grain curvature $K_{\mathrm{g}}$ and the bond curvature $K_{\mathrm{b}}$. However, near the grain/bond interface (see Figure \ref{fig:meta_geom}, around $\hat{x}^{*}$), the curvature is a mix of these two profiles \citep{Swinkels1981, Molaro2019}.
Additionally, in reality, shape irregularities can induce local curvatures that influence sintering. To account for such irregular shapes, rather than considering separate curvatures for the grain and bond, the model could be expanded to incorporate a continuous curvature function along the surface $\mathcal{S}$ of the system .
For instance, in Figure \ref{fig:meta_geom}, $r(\hat{x})$ describes the system's shape along the y-axis, and by calculating the second derivatives of $r$, one could derive the local curvature.

A similar approach can be found in the numerical scheme proposed by \citep{Flin2003} for irregular 3D-shaped grains. 
At each iteration, the surface normal of every voxel is calculated to determine its local curvature, which is then used to compute the local Hertz-Knudsen flux. 
In our model, such a scenario would lead to the modification of the gas mass variations (Equation \ref{eq:dmdt}) to:
\begin{equation}
	\dfrac{\partial }{\partial t} m_{\mathrm{gas}} = \int_{\hat{x} \in \mathcal{S}} J(\hat{x})  \, \mathrm{d}S(\hat{x})
	\label{eq:dmdt_irregular}
\end{equation}
where $\hat{x}$ are the points along the surface $\mathcal{S}$.
While such a precise formulation is essential for terrestrial studies, its relevance for icy moons is limited due to our lack of information about their current microstructure.
For Europa, even constraining grain sizes is challenging, so estimating the potential shape of the grains would be even more difficult.
Additionally, calculating surface normals and local curvatures, even in a 2D model, would add significant computational overhead. Given the need to compute sintering on a million years timescale and our current computational resources, implementing such a model would be very challenging.

\subsubsection{Grain size distribution}

Our current model of sintering only considers a uniform distribution of grain sizes, leading to bond growth at the expense of the grains. Experimental observations show that in granular ice media with a nonuniform grain distribution, grain growth and mass redistribution occur due to grain absorption \citep{Kuroiwa1961,  Molaro2019}. Smaller grains, with higher radii of curvature, experience a higher sublimation flux. As grains become smaller, sublimation rapidly removes mass at an increasing rate. Consequently, smaller grains tend to be absorbed by larger ones, leading to an increase in the mean grain size over time. This complex behavior is not captured by the model's idealized uniform grain size distribution.

However, while the grain radius decreases in our current model, the newly formed grain-and-bond system optically appears as a larger agglomerate structure.  A light ray penetrating a sintered medium may traverse through multiple grains before reaching a pore space, depending on the bond size. As a consequence, these grain-and-bond systems would be interpreted as larger grains with an effectively increased radius in spectroscopy or reflectance models. The exact relationship between bond radius and effective optical grain size is complex, as it also depends on the bonds' orientation within the observed surface. Further work would be necessary to properly model the optical behavior of sintered particles in spectroscopy models.

\subsubsection{Alternative transport mechanisms}

In this study, we have looked at the sintering through vapor transport diffusion where grains can only exchange matter with their neighboring bonds.
While the thermal properties considered may vary, the heat transfer simulations presented in the joint publication \citep{Mergny2024h} reveal temperature fluctuations of up to $\SI{15}{K}$ within the first $\SI{10}{cm}$.
This exceeds the $\SI{10}{K.m^{-1}}$ threshold, beyond which temperature gradient sintering becomes significant in Earth-based studies \citep{Colbeck1983}.
However, it remains unclear how much this threshold is modified in the cold, near-vacuum conditions found on the surfaces of icy moons.
Temperature gradients sintering could potentially play a crucial role in the surfaces of icy moons, and it an aspect that has yet to be undertaken by the planetary science community.
In this study, we have deliberately restricted pore communication to not account for this effect. 
Yet, Europa's surface is expected to be highly porous, with ice made of open pores that may have a vapor pressure gradient with depth. 
The main difference with terrestrial modeling is that the top layer is not in contact with the atmosphere but with the vacuum - more precisely, the near vacuum pressure of the exosphere. 
Given these factors, temperature gradients could play a significant role in the sintering of ice grains but its implementation would require additional modeling work.

\subsubsection{Electron-induced sintering}
The surface of Saturnian satellites undergoes radiation-induced sintering \citep{Schaible2017}, which may also affect Europa, as this process is closely linked to the distribution of high-energy electrons ($>\SI{}{MeV}$).
The flux of high-energy electrons is higher on Europa than on Saturnian moons \citet{Cooper2001, Schaible2017}, suggesting radiation-induced sintering likely occurs on Europa as well. However, the high surface temperatures of Europa, especially near the equator where most high-energy electrons impact \citep{Paranicas2009}, suggest that sintering is likely dominated by temperature rather than electron radiation.
It remains unclear if this process is significant for colder regions and if electrons below $\SI{}{MeV}$ can contribute to grain sintering.
The most notable difference for Europa is that grain sizes are expected to be much larger than those on Saturnian satellites. For radiation-induced sintering to occur in $\sim\SI{30}{kyrs}$ on Mimas and Tethys, grain sizes need to be smaller than $\SI{5}{\micro m}$, while for $\SI{25}{\micro m}$ grains, it takes $\SI{10}{Myr}$ \citep{Schaible2017}. Near-infrared spectroscopic measurements on Europa \citep{Hansen2004, CruzMermy2023} indicate that most models require grain sizes $>\SI{20}{\micro m}$. Therefore, over the million-year simulation time, at comparable plasma flux, such sintering is unlikely to be significant, even in polar regions, unless the grain sizes are much smaller than expected.
However, due to the higher electron flux on Europa and the likely non-uniform grain size distribution, smaller grains of a few microns could be affected by electron-induced sintering. Detailed modeling of such interactions would be beneficial but is beyond the scope of this paper and may be addressed in future versions of LunaIcy.

\section{Conclusion and Perspectives}

We have developed a multiphysics simulation model to study the evolution of planetary ice microstructure. 
Our approach includes MultiHeaTS an efficient heat transfer solver, as described in our joint paper \citep{Mergny2024h}, particularly adapted for scenarios with long timescales and small timesteps. Solar flux at the surface is calculated using Europa's orbital parameters.
Coupled to the heat transfer, we have built a new model, called LunaIcy, for the sintering of ice grain, that is based on the literature but with refinement of the diffusion process.
As changes in ice microstructure affect the thermal properties we have expressed the heat conductivity with a formulation that consider microstructure and porosity which enables a two-way coupling between sintering and heat transfer.

Our simulations spanned a million years, allowing us to thoroughly explore the evolution of Europa's icy surface microstructure. Results show that the hottest regions experience significant sintering, even if high temperatures are only reached during a brief portion of the day. This process takes place on timescales shorter than Europa's ice crust age, suggesting that these regions should currently have surface ice composed of interconnected grains.
For further studies, it would be valuable to extend this simulation time to the age of Europa's ice crust, estimated to be around $\SI{30}{My}$ \citep{Pappalardo1998}, but it would necessitate further optimization and available resources for computation. 

Other parameters describing the ice microstructure, such as crystallinity, roughness or porosity, may evolve due to processes such as space weathering, compaction or crystallisation. In a future work, we would like to expand LunaIcy by including more physical processes. Just as General Circulation Models (GCMs) have become essential tools for studying the climate evolution of planetary atmospheres, we anticipate further development of such multiphysics simulation tools for studying planetary surfaces.

Accurately simulating these highly coupled processes, such as sintering, provides valuable insights into the evolution of Europa's ice microstructure. These insights, in turn, contribute to the refinement of surface measurements like spectroscopy, enabling improved constraints on grain size. Such advancements play a crucial role in accurately determining the microstructure and quantitative composition of Europa's surface, a key objective for upcoming missions such as JUICE and Europa Clipper, and lander/cryobot missions \citep{Pappalardo2013, Hand2022, ValePereira2023} in a more distant future.

\appendix

\section{Geometry of the Sintering Model}
\label{sec:app_geom}
The surface areas in contact with phase exchange are determined using geometrical considerations based on the 1/4 grain-and-bond system of Figure \ref{fig:meta_geom}.

\subsection{One grain connected by two bonds} \label{sec:app_twobonds}

The main scenario investigated in this article is the case of two bonds per grain where the total volume is obtained by multiplying the surfaces and volumes of the 1/4 system shown in Figure \ref{fig:meta_geom} by a factor 4.
To begin, we consider the grain surface which is in contact with the pore space. The grain's surface is given by the integral
\begin{equation}
	S_{\mathrm{g}} = 4  \int_{0}^{\theta^{*}} \int_{0}^{\pi} (r_{\mathrm{g}} \,  d\theta) (r_{\mathrm{g}} \cos{\theta} \, d\Psi) 
\end{equation}
where $\Psi$ is the angle made by rotation along the $\hat{y} \hat{z}$ plane as shown in Figure \ref{fig:meta_geom}. 
After performing the integration, we arrive at the simplified expression
\begin{equation}
	S_{\mathrm{g}} = 4 \pi r_{\mathrm{g}} \hat{x}^{*}
    \label{eq:app_grain_surf}
\end{equation}

Next, we consider the bond surface, which is in contact with the pore space. The bond's surface is given by the integral:
\begin{equation}
	S_{\mathrm{b}} = 4 \int_{{0}}^{{\pi}} \int_{{0}}^{{\phi_{\mathrm{\rho}}^{*}}}  (r(\phi_{\mathrm{\rho}}) \: d{\Psi}) (r_{\mathrm{\rho}} \: d\phi_{\mathrm{\rho}} ) .
	\label{eq:surf_bond}
\end{equation}
where $r(\phi_{\mathrm{\rho}}) = r_{\mathrm{\rho}}  + r_{\mathrm{b}} - r_{\mathrm{\rho}}  \cos \phi_{\mathrm{\rho}}$,  is obtained from trigonometry (see Figure \ref{fig:meta_geom}). After integration, this leads to the expression
\begin{equation}
	S_{\mathrm{b}} = 4 \pi r_{\mathrm{\rho}}  \left( (r_{\mathrm{\rho}}  + r_{\mathrm{b}} ) \phi_{\mathrm{\rho}}^{*} - r_{\mathrm{\rho}}  \sin \phi_{\mathrm{\rho}}^{*} \right),
 \label{eq:app_bond_surf}
\end{equation}
knowing that $\phi_{\mathrm{\rho}}^{*} = \theta^{*}$.

Similarly, we can derive the analytical relationship between the grain and bond radii and their respective volumes $v(r_{\mathrm{g}}, r_{\mathrm{b}})$.
The grain volume is defined by the integral:
\begin{equation}
	v_{\mathrm{g}} = 4 \left( \frac{1}{2} \int_{{0}}^{{\hat{x}^{*}}} {\pi r(\hat{x})^2} \: d{\hat{x}} \right).
\end{equation}
By introducing a change of variable to the angle $\theta$ given by
\begin{equation}
	\begin{cases}
	  \hat{x} &= r_{\mathrm{g}} \sin \theta \\
	d\hat{x} &=  r_{\mathrm{g}} \cos \theta \: d\theta
	\end{cases}
	\label{eq:changextheta}
\end{equation}
and knowing that $r(\hat{x}) = r_{\mathrm{g}} \cos \theta$, we can express the integral as:
\begin{equation}
	v_{\mathrm{g}} = 2 \int_{{0}}^{{\theta^{*}}} {\pi r_{\mathrm{g}}^3 \cos^3 \theta} \: d{\theta} {}
\end{equation}
After integration over $\theta$, we get the following expression of the grain volume
\begin{equation}
	v_{\mathrm{g}} = \frac{\pi}{6} r_{\mathrm{g}}^3 \left( 9 \sin \theta^{*} +  \sin 3 \theta^{*} \right).
	\label{eq:app_grain_vol}
\end{equation}

Next, we consider the bond volume, which is defined by the integral:
\begin{equation}
	v_{\mathrm{b}} = 4 \left( \frac{1}{2} \int_{{\hat{x}^{*}}}^{{r_{\mathrm{g}}}} {\pi r(\hat{x})^2} \: d{\hat{x}} \right).
\end{equation}
By introducing a change of variable
\begin{equation}
	\begin{cases}
	 \hat{x} &= r_{\mathrm{g}} - r_{\mathrm{\rho}}  \sin \phi_{\mathrm{\rho}} \\
	d\hat{x} &=  - r_{\mathrm{\rho}}  \cos \phi_{\mathrm{\rho}} \: d\phi_{\mathrm{\rho}}  \\
	\end{cases}
	\label{eq:changevarsurf}
\end{equation}
we can express the integral as
\begin{equation}
	v_{\mathrm{b}} = 2 \pi \int_{{0}}^{{\phi_{\mathrm{\rho}}^{*}}} {\left( r_{\mathrm{\rho}}  + r_{\mathrm{b}} - r_{\mathrm{\rho}}  \cos\phi_{\mathrm{\rho}} \right)^2 }  r_{\mathrm{\rho}}  \cos \phi_{\mathrm{\rho}}\: d{\phi_{\mathrm{\rho}}}.
	\label{eq:vol_bond}
\end{equation}
after integration it leads to the expression of the bond volume
\begin{equation}
	v_{\mathrm{b}} = 2 \pi  r_{\mathrm{\rho}}   \left[\sin \phi_{\mathrm{\rho}} \left( (r_{\mathrm{\rho}} +r_{\mathrm{b}})^2 + \frac{9}{12} r_{\mathrm{\rho}}^2 \right)   
							   - \frac{1}{2} \sin 2\phi_{\mathrm{\rho}} \left( r_{\mathrm{\rho}}^2 + r_{\mathrm{\rho}} r_{\mathrm{b}}  \right) 
				  + \sin 3\phi_{\mathrm{\rho}} \, \frac{r_{\mathrm{\rho}}^2}{12} 
				  -\phi_{\mathrm{\rho}} \left( r_{\mathrm{\rho}}^2 + r_{\mathrm{b}} r_{\mathrm{\rho}} \right)  \right] .
	\label{eq:app_bond_vol}
\end{equation}
Now that analytical expressions  of the grain and bond volumes were obtained, the Newton-Raphson method can be called to retrieve the bond and grain radii.
Finally, it is worth mentioning that the pore volume is not required by the algorithm, as the expression of the gas pressure in the steady state is independent of $ V_{\mathrm{p}}$. 
However, if one wants to estimate the timescale of gas mass variations $\tau_{\mathrm{gas}}$, the pore volume can be approximated, at first order, to
\begin{equation}
    V_{\mathrm{p}} = \frac{4}{3} \pi r_{\mathrm{\rho}}^3.
\end{equation}

\subsection{One grain connected by a single bond} 

For a grain with a single bond, the total surface and volume of the bonds are simply divided by 2 from Equation \ref{eq:app_bond_surf} and Equation \ref{eq:app_bond_vol}.
The total grain surface is obtained by adding half the surface of a sphere to half the surface of the two-bonds case (see Equation \ref{eq:app_grain_surf}):
\begin{equation}
	S_{\mathrm{g}} = 2 \pi r_{\mathrm{g}}^{2} + 2 \pi r_{\mathrm{g}} \hat{x}^{*}.
    \label{eq:app_grain_surf_1bond}
\end{equation}
Similarly, the total grain volume is found by adding half the volume of a sphere to half the volume calculated in the two-bonds case  (see Equation \ref{eq:app_grain_vol}):
\begin{equation}
	v_{\mathrm{g}} = \frac{2}{3} \pi r_{\mathrm{g}}^3 + \frac{\pi}{12} r_{\mathrm{g}}^3 \left( 9 \sin \theta^{*} +  \sin 3 \theta^{*} \right).
	\label{eq:app_grain_vol_1bond}
\end{equation}

\subsection{One grain connected by six bonds} 

For a grain with six bonds, the total surface area and volume of the bonds are simply multiplied by 3 from Equation \ref{eq:app_bond_surf} and Equation \ref{eq:app_bond_vol}.
The total grain surface is calculated by subtracting the areas of the six caps removed by the bonds from the area of a sphere. The surface of a single cap is given by
\begin{equation}
	S_{\mathrm{cap}} = \int_{0}^{2\pi} \int_{\theta^{*}}^{\frac{\pi}{2}} (r_{\mathrm{g}} \,  d\theta) (r_{\mathrm{g}} \cos{\theta} \, d\Psi) = 2 \pi r_{\mathrm{g}} ( r_{\mathrm{g}} - \hat{x}^{*})
\end{equation}
resulting in the total grain surface of 
\begin{equation}
	S_{\mathrm{g}} = 4 \pi r_{\mathrm{g}}^2  - 12 \pi r_{\mathrm{g}} (r_{\mathrm{g}} - \hat{x}^{*}).
    \label{eq:app_grain_surf_6bond}
\end{equation}
The total grain volume is calculated by subtracting the volume of the six caps removed by the bonds from the volume of a sphere. The volume of a single cap is given by
\begin{equation}
	v_{\mathrm{cap}} =  \int_{\hat{x}^{*}}^{r_{\mathrm{g}}} {\pi r(\hat{x})^2} \: d{\hat{x}} = \frac{\pi}{12} r_{\mathrm{g}}^3 \left( 8 - 9 \sin\theta^{*}  - \sin 3\theta^{*} \right) 
\end{equation}
resulting in the total grain volume of 
\begin{equation}
	v_{\mathrm{g}} = \frac{4}{3} \pi r_{\mathrm{g}}^3 -  \frac{\pi}{2} r_{\mathrm{g}}^3 \left( 8 - 9 \sin\theta^{*}  - \sin 3\theta^{*} \right) 
	\label{eq:app_grain_vol_6bond}.
\end{equation}

\section{Steady state fluxes} \label{sec:app_ssflux}

Using Equation \ref{eq:mgas_infty} and the ideal gas law (Equation \ref{eq:perfect_gas}), the gas pressure at the equilibrium simplifies to
\begin{equation}
    P_{\mathrm{\infty}} = P_{\mathrm{s}}(T) \left[ 1 + \frac{\gamma M}{S R T \rho_{\mathrm{0}} }\left( S_{\mathrm{b}} K_{\mathrm{b}} + S_{\mathrm{g}} K_{\mathrm{g}} \right)  \right].
\end{equation}
By recognizing the limited development of the exponential function
\begin{equation}
    P_{\mathrm{\infty}}  = P_{\mathrm{s}}(T) \exp \left( \frac{\gamma M}{S R T \rho_{\mathrm{0}} }\left( S_{\mathrm{b}} K_{\mathrm{b}} + S_{\mathrm{g}} K_{\mathrm{g}} \right)  \right) 
\end{equation}
we identify the expression of the curvature pressures as in Equation \ref{eq:Kelvin}, which leads to a simpler expression of $P_{\mathrm{\infty}}$
\begin{equation}
     P_{\mathrm{\infty}} = \left( P_{K_{\mathrm{b}}}^{S_{\mathrm{b}}} P_{K_{\mathrm{g}}}^{S_{\mathrm{g}}} \right)^{1/S}.
\end{equation}
This expression clearly shows that at the steady state, the gas pressure is not the saturated vapor pressure.

Injecting the gas pressure at the steady state into equation Equation \ref{eq:Hertz-Knudsen}, we obtain the surface flux at the steady state
\begin{equation}
    J_{\mathrm{j}}  = P_{\mathrm{s}}(T) \cfrac{\alpha\gamma}{\sqrt{2 \pi } S  \rho_{\mathrm{0}}}  \left( \dfrac{M}{RT} \right)^{\frac{3}{2}} \left( K_{\mathrm{j}} S - K_{\mathrm{b}} S_{\mathrm{b}} - K_{\mathrm{g}} S_{\mathrm{g}}\right) 
\end{equation}
Due to conservation of mass, at the steady-state Equation \eqref{eq:dmdt} shows that the total mass sublimating from the grains should be equal to the total mass condensing on the bonds:
\begin{equation}
    J_{\mathrm{b}} S_{\mathrm{b}} + J_{\mathrm{g}} S_{\mathrm{g}} = 0
    \label{eq:dmdt_steady}
\end{equation}
So we can first compute the surface mass flux entering the pore space from the bonds (which is negative)
\begin{equation}
    J_{\mathrm{b}}   = P_{\mathrm{s}}(T)  \cfrac{\alpha\gamma}{\sqrt{2 \pi } S  \rho_{\mathrm{0}}}  \left( \dfrac{M}{RT} \right)^{\frac{3}{2}} S_{\mathrm{g}} \left( K_{\mathrm{b}} - K_{\mathrm{g}} \right)
\end{equation}
and the grain flux is simply retrieved using Equation \ref{eq:dmdt_steady}.

\newpage
\section*{Statements and Declarations}
\subsection*{Funding and Competing Interests}

We acknowledge support from the ``Institut National des Sciences de l'Univers'' (INSU), the ``Centre National de la Recherche Scientifique'' (CNRS) and ``Centre National d'Etudes Spatiales'' (CNES) through the ``Programme National de Plan{\'e}tologie''. 

The authors have no competing interests to declare that are relevant to the content of this article.

\section*{Data Availability}
The authors assert that the data supporting the study's findings are included in the paper, while some of the code is accessible on the online public repository at the IPSL Data Catalog:\dataset[10.14768/9763d466-db02-4f29-8ad5-16e6e0187bd4]{\doi{10.14768/9763d466-db02-4f29-8ad5-16e6e0187bd4}}, with the remainder available upon request.

%\newpage
%\section*{Figures}

\bibliography{library}
\bibliographystyle{aasjournal}

\end{document}